\documentclass[graybox]{svmult}

\usepackage{type1cm}        
\usepackage{makeidx}         
\usepackage{graphicx}        
\usepackage{multicol}        
\usepackage[bottom]{footmisc}

\usepackage{xcolor}
\usepackage{newtxtext}       %
\usepackage{newtxmath}       
\usepackage{longtable}
\usepackage{booktabs}
\usepackage{microtype}

\usepackage[T1]{fontenc}
\usepackage[utf8]{inputenc}

\usepackage{hyperref}
\hypersetup{
 colorlinks=true,   
 citecolor=black,
 urlcolor=black,
 linkcolor=black,
 bookmarksnumbered=true,     
 bookmarksopen=true,         
 bookmarksopenlevel=1,          
 pdfstartview=Fit,           
 pdfpagemode=UseOutlines,    
 pdfpagelayout=TwoPageRight 
}

\renewcommand{\tabcolsep}{0.15cm}

\makeindex             

\begin{document}

\title*{Teaching Theorizing in Software Engineering Research}
\titlerunning{Teaching Theorizing in Software Engineering Research} 

\author{Klaas-Jan Stol}
\institute{Klaas-Jan Stol \at Lero and University College Cork, Western Road, Cork, Ireland \email{k.stol@ucc.ie}}
\maketitle

\abstract{This chapter seeks to support software engineering (SE) researchers and educators in teaching the importance of theory as well as the theorizing process. Drawing on insights from other fields, the chapter presents 12 intermediate products of theorizing and what they mean in an SE context. These intermediate products serve different roles: some are theory products to frame research studies, some are theory generators, and others are components of a theory. Whereas the SE domain doesn't have many theories of its own, these intermediate products of theorizing can be found widely. The chapter aims to help readers to recognize these intermediate products, their role, and how they can help in the theorizing process within SE research. To illustrate their utility,  the chapter then applies the set of intermediate theorizing products to the software architecture research field. The chapter ends with a suggested structure for a 12-week course on theorizing in SE which can be readily adapted by educators.
}

\section{Introduction}
The ancient Greeks are widely acknowledged to have made major advances in the sciences \cite{nisbett2004geography}. Why might this be? Nisbett describes \cite[p. 4]{nisbett2004geography}:

\begin{quote}
The Greeks, far more than their contemporaries, speculated about the nature of the world they found themselves in and created models of it. They constructed these models by categorizing objects and events and generating rules about them that were sufficiently precise for systematic description and explanation. [...] Whereas many great contemporary civilizations, as well as the earlier Mesopotamian and Egyptian and the later Mayan civilizations, made systematic observations in all scientific domains, only the Greeks attempted to explain their observations in terms of underlying principles. 
\end{quote}

The success of the ancient Greeks thus lay in recognizing the value in creating models, categories, concepts, rules, and principles to establish an understanding of the world. 
How does research today fare in terms of these activities?
More recently than ancient Greek civilization, Suddaby \cite{suddaby2006editors} lamented that a common problem with Grounded Theory is ``a failure to `lift' data to a conceptual level,'' and indeed, this is also a shortcoming within SE research \cite{stol2016grounded}. 
Robson \cite{robson2002real} argued that \textit{``without theory the research may be easier and quicker, but the outcome will often be of little value.''} 

The last statement should hit a nerve among SE researchers. The SE community produces thousands of research papers each year, many of which have gone through a tedious review process and have found a willing audience among fellow researchers, and yet their value appears to be of little relevance or interest to SE professionals \cite{beecham2013we,weyuker2011empirical}. At the same time, a lack of attention for theory and conceptualization has been observed within the SE community \cite{johnson2012s,stol2015theory}. 
I argue that these two observations, namely (1) that much of SE research is not valued by practitioners, and (2) SE research lacks attention for theory and conceptualization, are correlated.
Perhaps, as a community, we should learn a lesson from the ancient Greeks, and focus on explaining our observations beyond the immediate settings we study, and on identifying underlying principles of our findings. This chapter seeks to draw attention to increase the reader's sensitivity to not only the need to theorize, but also to suggest a light-weight approach to get started on this endeavor. Or, with apologies to Larry Wall:\footnote{Perl creator Larry Wall's 11th State of the Onion speech was entitled: ``Programming is Hard, Let's go Scripting...'' \url{https://www.perl.com/pub/2007/12/06/soto-11.html/}} ``Theory development is hard, Let's go theorizing...''

The SE community has made major strides in improving researchers' level of expertise in using a wide range of research methods. However, conducting an excellent case study or a carefully designed experiment does not, on its own, add to the body of knowledge within SE. 
Whereas the remaining chapters in this volume focus on teaching specific research methods, we also need to consider the role of theory to advance science. Perhaps the single-most important aspect of any study is its motivation, its grounding in prior work, and how this feeds back to the literature. Theorizing can help to address this, because it encourages to build on or challenge prior work, and theorizing effectively can suggest promising research directions. Theory development is a fundamental research strategy \cite{stol2018abc}, but one that has not received much explicit attention. SE researchers simply aren't trained to theorize, and as a result, the publication norms within the SE community are such that nobody appears to miss it, either.

\subsection{The Need for Theorizing}

Why do we need theorizing? There are several reasons.
The first reason to engage in theorizing is that this distinguishes science apart from consulting or journalism. Presenting some findings that remain completely disconnected to other findings without drawing any parallels does not help a scientific field forward. 
In other words, without theory, there is no cumulative tradition, no `glue' that holds studies together, nor is there any hope for parsimony: tens or hundreds of studies might find similar findings, all slightly different, without any attempt to synthesize, generalize, and unify such findings into a coherent theory. Each study stands on its own, and while it may reference other studies on a particular topic, there is no attempt to provide any link at a conceptual level that helps us understand trends, causes, and effects.
Dubin distinguished between ``being a good reporter'' and ``being a good scientist''; the former can present new and useful information to the body of knowledge, but the latter is able to find relationships between `things' \cite{dubin1978theory}.
This can readily be seen in many SE papers: `Related Work' sections read as obligated shopping lists of references to please reviewers, rather than an attempt to synthesize the state of knowledge relevant to the study that is presented (``good reporting,'' not ``good science''). The common practice in SE papers of placing such a section \textit{at the end} of a paper exacerbates this, and is therefore misguided. 

Second, and closely related to the argument above, is that due to a lack of theory or conceptualization, we remain stuck at the level of the idiosyncratic, the specific, without seeing the `big picture.' Theory helps us to focus on what is important, rather than what is incidental. Without theory we tend to focus on the `here and now,' rather than on what we can learn to serve us in the future. That is, rather than focusing on the latest trends, which at the time of writing seems to be the use of Large Language Models (LLM) in Software Engineering (SE), and highlighting how useful specific features of a given version of ChatGPT might be, we ignore more general questions that consider fast-emerging technologies in general and how they can be leveraged in SE in practice. By the time a paper is published, the study may be outdated as the next version of the LLM is released! What can we learn from quickly emerging AI technology \textit{in general}? Or more generally stated, what can be a \textit{lasting contribution} to the field, rather than one that is outdated by the time the paper is presented at a conference? 

Third, theorizing can help to explain seemingly `obvious' findings. Many papers receive review comments stating that \textit{``the findings are obvious.''} The first line of defense against this is to state that, without an empirical study, we cannot know whether something is true; perhaps our assumptions are not true at all! Theorizing plays a key role here because it can help to unpack a line of reasoning, to lay bare the assumptions that may be held within the literature, and to make explicit, with rigorous and sound argumentation, what the underlying process might be that underpins the findings. If the data support the hypothesis, then besides presenting the first empirical evidence for it, a paper would also make explicit why those findings make sense. If findings are counter-intuitive, this by itself would also be very interesting. It is thus the unpacking of a given phenomenon or topic in detail---the theorizing---that would be the contribution.

\subsection{Overview of this Chapter}
This chapter draws on insights from other fields, in particular Information Systems (IS), to recast the discussion from ``theory in SE'' to ``theorizing in SE.''  
Specifically, this chapter is organized around 12 intermediate products of theorizing  identified by Hassan et al. \cite{hassan2022}. In previous work, I used the term ``theory fragments'' to refer to a similar idea \cite{stol2015theory}. I define ``theory development'' as a research strategy \cite{stol2018abc} that explicitly sets out to develop new theory. \textit{Theorizing} is the process of engaging  in theoretical development without the explicit goal of presenting a sufficiently developed theory.\footnote{Theories are arguably never complete, because they can always be extended in one way or another; hence, the phrase `sufficiently developed.'} Clearly, these activities are related; a scientific discipline needs both, but in this chapter I focus on the latter.

The goal of this chapter is twofold. First, it provides a gentle introduction to the 12 theorizing products, what they mean in general, and what they mean in a SE research context. The chapter then demonstrates how some excellent SE scholars have used these theorizing products. 
It is worth noting that one could fill a chapter on each of the 12 theorizing products. Many authors have presented insightful accounts on topics such as myths and metaphors; the current chapter only scratches the surface. Studying each of these is left to the reader as the proverbial exercise.
Second, this chapter encourages the reader to introduce conceptual thinking into their research. In what ways can a topic of interest be elevated to a more conceptual level? What theorizing products have been used in the reader's field of interest? How can the reader incorporate these in their own research? 

\begin{svgraybox}
\textbf{Learning Objectives}\\
The learning objectives of this chapter are:
\begin{enumerate}
    \item To understand and appreciate the role of theorizing in SE research.
    
    \item To understand and apply the 12 intermediate theorizing products in the context of SE research;
    \item To recognize theorizing in SE papers, by reconstructing the theoretical advances made using the 12 intermediate theorizing products.
\end{enumerate}
\end{svgraybox}

\section{Theory Development in Software Engineering Research}

\subsection{Views on Theory in Software Engineering Research}
Whereas there have been some efforts within the SE literature to pay more attention to `theory,' the SE community has, by and large, ignored these. One reason for this is that theory development is difficult. Another reason is that there appears to be a mismatch between what researchers \textit{believe} theory is, and the research that they do; researchers may believe that traditional notions of theory, such as a network of constructs and relationships among them, do not apply to their research. 

These are good reasons, and they have to do with the fact that nobody appears to be able to define what `theory' \textit{is}. SE researchers are not alone in this; it appears fair to say that there are no objective criteria by which one can determine what is, and what is not, a theory.

In this chapter I argue that, in order to learn how to theorize, we need not be overly concerned with the traditional notion of theory as a collection of constructs and relationships, but that intermediate theorizing products \cite{hassan2022} may also be sufficient to move the SE field forward. These outcomes can take many different shapes; the chapter seeks to explain the value of each of these types of products, what their added value is, and how they can contribute to a cumulative tradition in SE research.  The chapter seeks to `free' readers from thinking in fixed patterns (traditional notions of `theory') and offer new ways that might be more suitable to SE.

Researchers in the SE field have varying views on theory and its role in research. This is partly due to the fact that the SE field consists of several sub-fields, many of which have their own dedicated conferences and sometimes journals, and consequently, each with different ideas and norms about what constitutes `research.' For example, the MSR\footnote{Mining Software Repositories} community is quite distinct from the CHASE\footnote{Cooperative and Human Aspects in Software Engineering} community in the way the research is done. Some researchers have looked, seemingly with a sense of envy, at other academic disciplines where it is more common to have theories that guide and inform empirical studies, which in turn extend or dispute those theories. This envy has led some scholars to wonder: \textit{where is the general theory of software engineering?} 
\cite{johnson2012s}.

Others do not seem to perceive this lack of theoretical focus, and may consider the state of the art within their specific sub-field as constituting the theory that they contribute to. If we were to ask researchers active in requirements engineering or software testing, for example, to provide an overview of their theory, we'd get models, processes, and taxonomies (cf. \cite{bertolino2007software,pohl1996requirements}). The term `theory' is used here in a different way, yet it is not fundamentally incompatible with how it is used in the social sciences. In fact, the intermediate theorizing products presented in this chapter can be identified in those bodies of knowledge as well. Doing so is a challenging but excellent exercise for the reader.

\subsection{What Do Theories Look Like?}\label{sec:variance_process_theory}
\begin{figure}[!ht]
    \centering    
    \includegraphics[width=0.8  \textwidth]{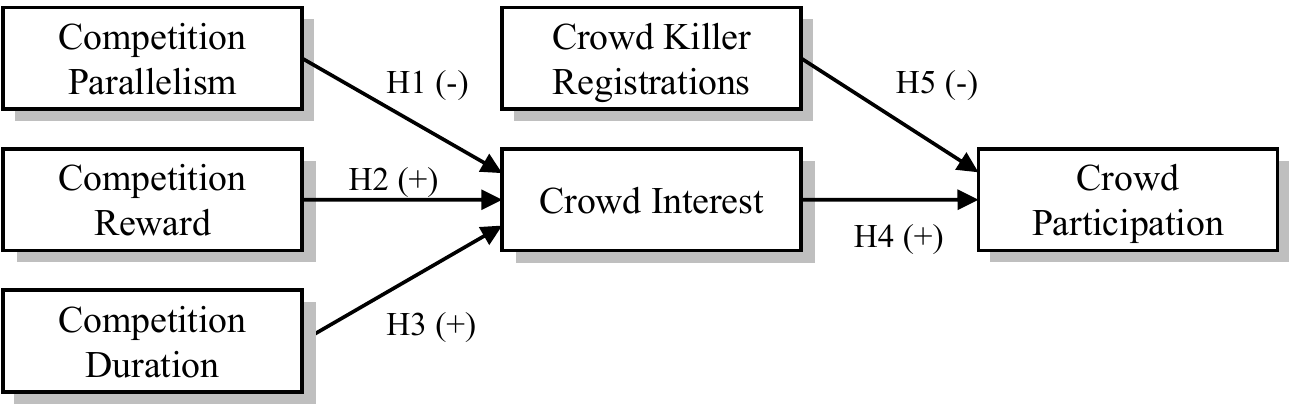}    
    \caption{Theory of crowd participation \cite{stol2019competition}, an example of a variance theory}
    \label{fig:variancetheory}
\end{figure}

Most discussions of `theory' in SE provide tutorial-style explanations suggesting that a theory consists of theoretical concepts and the relationships between them (propositions or hypotheses) (cf. \cite{stol2015theory}). This common format is a template for what is called a variance theory: if A causes B, then A and B are the concepts, and the relationship is a causal one.\footnote{Most work tends to shy away from the verb `cause' and use the more vaguely stated ``is associated with'' so as to not claim causality. Causality can typically not be demonstrated without conducting controlled experiments (see also Vegas and Juristo's chapter on experimentation in this volume), though there are some techniques that can provide convincing evidence for causality.} This means that if A goes up, then B goes up also. Figure~\ref{fig:variancetheory} presents an example of a variance theory.  
Variance theories are intuitive, and they can be tested using quantitative datasets, which are relatively easy to come by, for example by mining software repositories (see Codabux et al.'s chapter in this volume) or developer surveys, notwithstanding the many challenges in both MSR and conducting surveys (see Kalinowski et al.'s chapter in this volume). 

\begin{figure}[!h]
    \centering        
    \includegraphics[width=\textwidth]{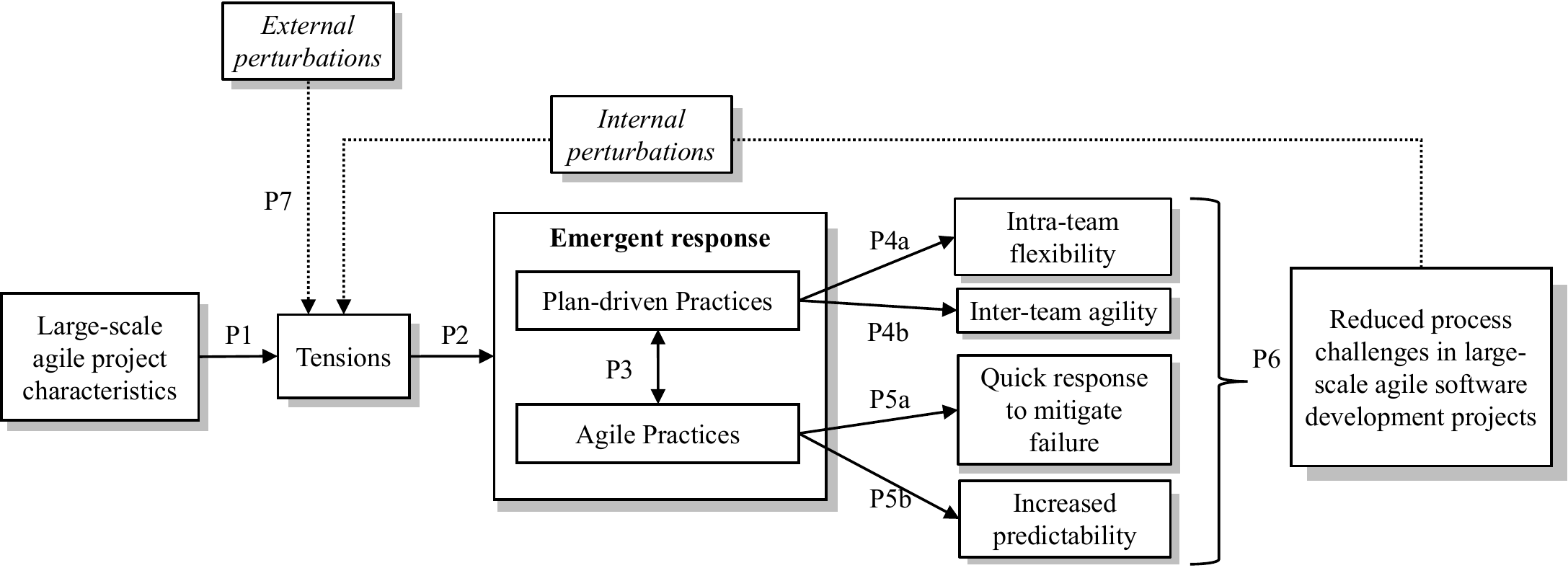}
    \caption{Process theory of large-scale Agile software development (Rolland et al. \cite{rolland2023}) }
    \label{fig:processtheory}
\end{figure}

Another type of theory is called `process theory,' defined by Huber and Van de Ven \cite{huber1995} as:
\textit{``An explanation of the temporal order in which a discrete set of events occurred, based on a story or historical narrative.''}
Ralph defines a process theory as: \textit{``A process theory is a system of ideas intended to explain, understand (and sometimes predict) how an entity changes and develops,''} and discusses four `ideal types' of process theories frequently found in organizational/management research, and how they apply to SE \cite{ralph2018toward}. 
Process theories are frequently more flexible and capture a wide range of phenomena in the real world. However, these theories are not easily tested using quantitative datasets only, but rather in comparison to an alternative process theory (cf. Ralph \cite{ralph2016software}). Figure~\ref{fig:processtheory}  presents 
a process theory of large-scale agile software development. This process theory explains how  tensions that arise through both internal and external perturbations require an emergent response of the organization or teams involved, so as to restore a certain `balance.' 
As such, this can be seen as a \textit{dialectic process theory} (one of the four `ideal' types), in which \textit{``stability and change are explained by reference to the balance of power between opposing entities''} \cite{van1995explaining}. In this case, the opposing entities are the tensions that arise due to a conflict between agile processes and the large-scale nature of a software project, and the emergent responses by means of adopting both plan-driven and novel agile practices that can reduce those tensions. Note that this process theory defines a number of propositions (numbered P1 to P7), and may resemble the hypotheses in the variance theory in Fig.~\ref{fig:variancetheory}, but that these are not easily testable using quantitative datasets.

While variance and process theories are perhaps the best known within SE literature, these are by no means the only forms. Other scientific disciplines and discourses follow other traditions. For example, Dittrich developed a \textit{practice theory} on the use of software development methods \cite{dittrich2016does}, which:
\begin{quote}    
``develops a conceptual base for understanding software development as social and epistemic practices, and defines methods as practice patterns that need to be related to, and integrated in, an existing development practice.''
\end{quote}
The format of Dittrich's presentation does not follow that of a variance or process theory, but is much more narrative. Thus, there is no clear-cut format that defines what theory is. This remains one of the major challenges in developing theory.

\subsection{Challenges in Developing Theory in Software Engineering Research}
\label{sec:challenges}

As noted earlier, while much has been written over the past decades on `theory,' there is no general consensus on a definition of `theory' \cite{wieringa2015six}, and what people accept as a theory appears to be dependent on the norms of a specific field of research. What might some other reasons be why developing theory is difficult in the SE field?

\begin{itemize}
    \item There is a general lack of knowledge and appreciation for theorizing;
    \item There is a mismatch between the SE field's fast pace of development and general level of patience of researchers;
    \item Researchers conflate and lack understanding of terminology.
\end{itemize}

Due to a general lack of knowledge of, and appreciation for theorizing, many researchers within the SE field simply do not pay much attention to this. Related work sections are mere shopping lists of a ``who's who'' in the field of interest, without a deep engagement with the actual contributions of those authors; Nobel Prize recipient Sir Peter Medawar described it as \cite{medawar1990scientific}:

\begin{quote}
    ``a section called `previous work' in which you concede, more or less graciously, that others have dimly groped towards the fundamental truths that you are now about to expound.''
\end{quote}

In other words, the SE field lacks an effective cumulative tradition. For example, there are over 100 case studies on the use of agile methods in distributed development \cite{desouza2024csur}.
Through a carefully crafted introduction, many papers over-rely on `spotting the gap,' i.e., identifying the study that \textit{hasn't} been done, without aiming to see the forest for the trees.

Another major impediment to developing theory is the time commitment that is necessary. Two related factors play a role which go against this time commitment. First, there is the SE field's pace: new technologies and trends emerge quickly. At the time of writing, the release of ChatGPT, a publicly available LLM, immediately demanded attention from many researchers leading to papers that describe this new type of technology. Second, SE researchers are not rewarded for long-term investment; it is still common for researchers to target the International Conference on Software Engineering (ICSE) with papers that are not further developed.\footnote{Without wishing to sound cynical or overly critical, I am not the first to lament this \cite{offutt2011purpose,parnas2007stop,briand2012embracing}.} It appears that the community norm isn't such that researchers are encouraged or expected to follow up, nor to develop theory. This community norm is difficult to change. Students may not be overly concerned with this, as their funding clock is ticking. Supervisors are in a difficult position, because they prefer their students to succeed and not subject them to higher standards than would be the norm within the community, while both funding agencies and tenure evaluation committees are expecting results from principal investigators.

Finally, the terminology surrounding theory, theorizing, and theory development can be rather confusing. Most of the terms discussed in this chapter are misunderstood, misused, or conflated with other terms. For example, the terms `model' and `framework' are frequently conflated, as are 
the terms `concept' and `construct' \cite{Sjoberg2008}. 

The remainder of this chapter addresses these three issues. By elaborating each of the 12 intermediate products of theorizing in the next three sections, it addresses the last issue of terminology. It then highlights 12 intermediate products of theorizing and demonstrates how they have appeared in SE literature, drawing attention to the role these products have played in theorizing. Finally, it draws attention to the possibility of building a cumulative tradition within SE, by demonstrating how the theorizing products can contribute to a `bigger picture.'

\section{Theory Framing}
The first set of theorizing products help in the activity of \textit{framing} a phenomenon. The term `framing' means establishing what the topic of interest is, or positioning a research problem in the real world. Framing is the activity of finding a `hook,' or problematizing prior work \cite{alvesson2011generating,rolland2023} to start off the theorizing activity. This section discusses theory products to frame studies with a summary in Table~\ref{tab:framing}.

\subsection{Question}
A starting point to frame or position a topic or phenomenon of interest is to ask questions. Hassan et al. \cite{hassan2022} pointed out that \textit{``a discipline is defined by the set of questions it asks.''} 
Different disciplines can ask related questions pertaining to the same phenomenon, but depending on the discipline, different concepts, methods, and tools are likely to be used to answer those questions. 
A good example to illustrate this is the question: \textit{What is Open Source Software?} 
OSS emerged in the nineties as a topic of interest for several disciplines. Economists were baffled by the phenomenon that developers would work \textit{for free} to deliver high quality products \cite{bonaccorsi2003open,lerner2002some}. Information Systems (IS) researchers tended to focus on OSS as a product, its qualities \cite{payne2002security,stamelos2002code}, and how organizations could leverage open source communities to build products \cite{aagerfalk2008outsourcing,morgan2014beyond}.
Meanwhile, SE researchers initially focused on the development process, asking questions surrounding concepts that were of particular interest to software engineers \cite{mockus2002two}. More recently, SE researchers have focused on a wide range of other aspects of OSS, such as peer review practices \cite{rigby2014peer} and corporate participation in OSS projects \cite{zhang2022corporate}.

How is a question an intermediate product of theory? Asking a question goes together with a justification, a motivation, and a perspective that is linked to a discipline, and knowledge of a field of inquiry. Doctoral students frequently spend a considerable amount of time to develop a research proposal, and it would not be uncommon for this process to take over one year. Identifying the question to pursue is far from trivial, because questions must come with a line of reasoning---and the lack of a clear reasoning is frequently a justified reason to reject a manuscript. The ability to ask interesting questions relies on intimate knowledge of both the state of practice and the state of the art. Asking interesting questions thus helps to articulate what is yet unknown,  in a way that is relevant and important to a field of inquiry, suggesting that an answer will help the field as a whole forward. Alternatively, questions can be asked by challenging what is supposedly already known: problematizing assumptions within a given body of literature can be a fruitful approach to generate new insights \cite{rolland2023,ryan2023ccs, sandberg2011ways}.

In conclusion, a question is an intermediate product of theorizing, because it conveys not only the question itself, but also an awareness and familiarity of prior literature, as well as the state of a particular field of inquiry. The question itself does not constitute a theory, but if answered, can lead to a theory.

\subsection{Paradigm}\label{sec:paradigm}

Whereas questions define a research discipline, the type of questions that are being asked within a given field depends on the \textit{paradigm} that researchers within that discipline subscribe to.\footnote{\label{foot:paradigm}The term `paradigm' is considered by some as somewhat controversial due to Thomas Kuhn's use of the term in his influential book \textit{``The Structure of Scientific Revolutions''} (cf. Shapere \cite{shapere1971}). Notwithstanding, a paradigm is a way of looking at the world, and comes with a set of shared assumptions. For example, in the Object-Oriented (OO) programming paradigm, all `things' are modeled as `objects.' A paradigm is reflected in the way people think, write, and discuss about a topic of interest. The term is not limited to only refer to \textit{research paradigms}, such as (post-)positivism, interpretivism, and critical realism.} 
A good example here is Basili's pioneering work that has emphasized the role of experimentation in SE research \cite{basili1977software}. 
Basili \cite{basili1993experimental} wrote:

\begin{quote}
    There is a fair amount of research being conducted in software engineering, i.e., people are building technologies, methods, tools, etc. However, unlike in other disciplines, there has been very little research in the development of models of the various components of the discipline. The modeling research that does exist has centered on the software product, specifically mathematical models of the program function.
\end{quote}
Basili's key point here is that there had been too much focus on building tools and technologies, rather than the main components of software engineering as a discipline. He argued for a need to make \textit{theoretical models}, not (only) mathematical (formal) models of software products, though obviously the latter is a valid way of doing research. Basili continued:

\begin{quote}
    We need research that helps establish a scientific and engineering basis for software engineering. To this end, the research methodologies required involve the need to build, analyze and evaluate models of the software processes and products as well as various aspects of the environment in which the software is being built, e.g the people, the organization, etc. It is especially important to study the interactions of these models. The goal is to develop the conceptual scientific foundations of software engineering upon which future researchers can build. 
\end{quote}

Whereas experimental SE emphasizes quantitative methods and the need to develop a conceptual scientific foundation, a next step in the evolution of empirical methods was evidence-based software engineering (EBSE), modeled after evidence-based medicine (EBM) \cite{kitchenham2004evidence}.\footnote{Following Kitchenham et al. \cite{kitchenham2004evidence} and the emergent attention for theory in SE about a decade ago, I had suggested a new paradigm: Theory-Oriented Software Engineering \cite{stol2015theory}. It appears the paradigm has thus far not taken any root.} 
The EBSE paradigm has had a major impact on the SE field. An important part of EBSE is the Systematic Literature Review (SLR) method. This method is commonly used in medicine to collect all available evidence for a certain `intervention' (for example, a medical treatment). Similarly, the SLR was envisaged to be a useful tool to gather all evidence regarding specific practices and tools, and indeed, countless SLRs are published every year. This is a good example of how the paradigm shapes the way we answer research questions.\footnote{While the idea of the SLR is a good one, unfortunately many SLRs in the field lack synthesis and theorizing that could help readers to see the forest before the trees \cite{cruzes2011research}.}

We can also observe a paradigm shift in how the SE community sees itself. Whereas traditionally the SE field was seen by many scholars as an engineering field \cite{baragry2000phd,briand2012embracing,shaw1990prospects,wasserman1996toward}, there is an increasing consensus that SE is a socio-technical field \cite{stol2016grounded}, focusing on what has been termed Behavioral Software Engineering \cite{lenberg2015behavioral}, and that software has a major impact on society; for example, since 2015, ICSE features a Software Engineering in Society (SEIS) track. In the last decade or so, there has been increasingly attention for human aspects; more recently, studies focus on themes such as burnout, developer happiness \cite{graziotin2018happens}, and diversity and inclusion.

In sum, a paradigm is not a theory, but helps to frame a topic of interest, to direct our focus, and to inform research questions to answer. To follow a certain paradigm, a researcher adopts a specific focus on what to study, and how.

\subsection{Law}

A `law' is a generalizing statement that holds true under certain conditions, and which provides explanatory power \cite{endres2003handbook}. Newton's law of universal gravitation in physics, for example, explains why objects fall to the ground. The general law of supply and demand in economics explains when prices of products and services go up or down (with boundaries and conditions beyond which the law may not hold, e.g., economic monopolies, price elasticity). In both cases, specific conditions may affect the extent to which these laws hold.

There are many `laws' in SE \cite{endres2003handbook,sachslaws}, though what exactly makes something a law is unclear. Many of them appear to be generalized statements that are based on observations in practice, though it appears that these statements have held over time. 
One of these is Brooks' Law \cite{brooks1987essence}, which states that adding manpower to a late software project makes it later.
Several researchers have taken Brooks' Law to \textit{frame} their research \cite{adams2009coordination,hsia1999brooks}.
Hsia et al. \cite{hsia1999brooks}, for example, conducted a system dynamics simulation study to understand the implications of Brooks' Law. 
The `law' suggests a certain truism, but Hsia et al.'s simulation study suggests that \textit{``there is a time line T for each project that if enough manpower is added before T, the project still can finish before the scheduled deadline,''} challenging the universality of Brooks' Law. 

A second example of a law in SE is Conway's Law \cite{conway1968committees}, which states that organizations produce system designs that mirror their communication structures. 
Herbsleb and Grinter \cite{herbsleb1999splitting} used Conway's Law as a starting point for their study of distributed development, where good communication, in particular informal communication, is much more challenging to achieve. Conway's Law has been used as a way to frame several other studies \cite{bowman1998software,bird2010conway}.

Laws are not quite theories, but generalizations that can be used to frame a research study, to declare an initial set of theoretical expectations. The extent to which a law holds under alternative conditions can then be the focus of the study, either to confirm the law or refute it.

\subsection{Framework}
A framework captures the things a researcher is interested in; it describes the ``map of the territory'' that is of interest \cite{miles1994qualitative,schwarz2007understanding}.\footnote{In this chapter I refer to theoretical or conceptual frameworks, not to be confused with software development frameworks, which are software libraries that provide functionality and may impose an architectural structure of the software that is developed. For example,  Django and Flask are popular frameworks for developing Python web applications.} 
A framework has been described as \textit{``a less developed form of a theory''} \cite{rudestam1992surviving}.
A framework identifies the key concepts within a study; it may also capture how the concepts of interest relate to one another, but that is not necessary. Any concept that is not part of the framework is thus not part of the `map,' and thus not studied. In other words, a framework helps the researcher to focus on what is important. 
The conceptual framework is usually developed prior to any empirical work through a literature survey---a thorough review of literature that involves synthesis and theorizing is a study on its own \cite{stol2018abc}. As part of the empirical work, the researcher may decide to refine the conceptual framework.

\begin{figure}[!ht]
    \centering
    \includegraphics[width=\textwidth]{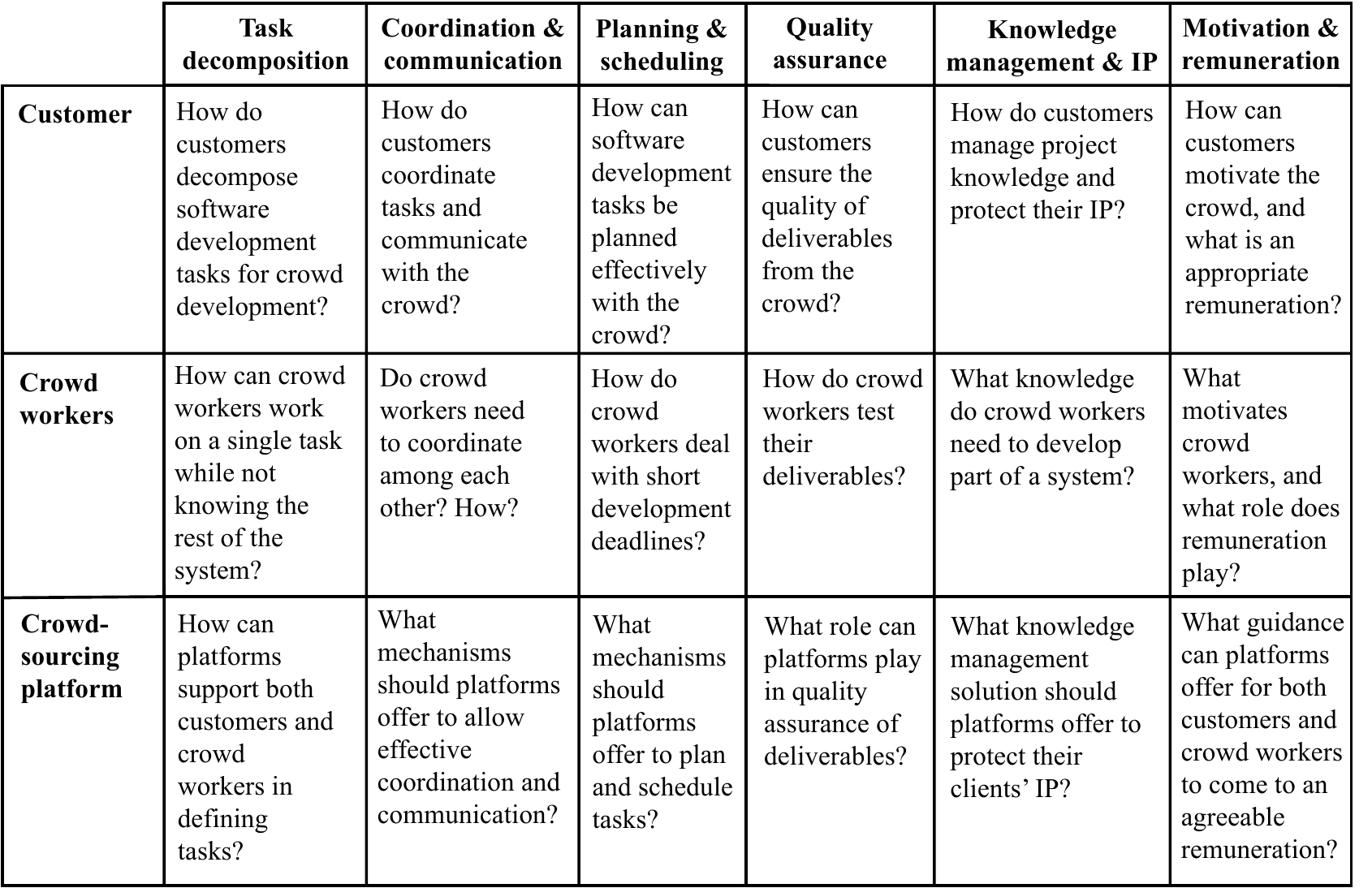}
    \caption{Conceptual framework capturing `facets' of crowdsourcing in a software development context, as well as three different perspectives from which these facets can be considered (based on Stol and Fitzgerald \cite{stol2014researching})}
    \label{fig:crowd}
\end{figure}

An example is Stol and Fitzgerald's framework to study crowdsourcing for software development \cite{stol2014researching,stol2014two}. 
The framework is developed based on an extensive survey of prior literature on crowdsourcing to develop an understanding of the process, what it entails, and who might be involved. The framework identifies six salient themes that appear to be of particular relevance to crowdsourcing in a software development context. Further, the crowdsourcing process involves different stakeholders or perspectives (see Figure~\ref{fig:crowd}).
This framework does not, however, capture any relationships between different concepts. 
Researchers can focus on one particular perspective across all six facets (one of the rows), or on one particular facet across three perspectives (one of the columns), or only on one facet from a single perspective (a single cell).

Developing the conceptual framework is a theorizing activity; there is no absolute right or wrong way of doing it, though good judgment and justification play a role. Which concepts and themes are included depends on the researcher who should justify each concept's inclusion, typically through a careful literature review. 
When developing a conceptual framework, one should ensure that the map doesn't become larger than the territory; that is, researchers may be overly diligent or mechanistic, and attempt to capture every little detail that could be relevant to their topic of interest. This would lead to a framework that is so complicated that it loses its strength; the point of a framework is to focus one's attention to the important elements within a domain. In my experience, frameworks that capture up to 7 themes are manageable; when more themes are included, it may be useful to add some level of hierarchy to organize concepts. For example, in our study of key factors for adopting Inner Source, we identified no fewer than nine factors, and so we organized these into three categories \cite{stol2014key}. 

In an attempt to keep the number of concepts limited, it can be tempting to combine concepts that appear to be related. It is important that this is done with utmost care, only when two concepts are so intertwined that keeping them separate is not practical. For example, several concepts within our framework for studying crowdsourcing were combined; when discussing the topic of `planning,' we could not clearly and meaningfully separate this from `scheduling.' Thus, if two concepts are intertwined, it is best to combine them. If there are too many distinct concepts, it is better to add some level of hierarchy by introducing categories. 

A framework is not a theory, but it certainly is an intermediate product. It helps to conceptualize important elements within a domain of interest and as such to \textit{frame} a research study. They are reusable and extendable by others, and can provide a foundation for a whole stream of studies. 

\begingroup
\footnotesize 
\begin{longtable}{p{1.9cm}p{4.9cm}p{4cm}}
\caption{Theory frames}\label{tab:framing}\\
    \toprule
    Product & Description & Example in SE\\
    \midrule
\endfirsthead
    \toprule 
    Product & Description & Example in SE\\
    \midrule
\endhead
   \bottomrule
\endfoot
    \bottomrule     
\endlastfoot

    Question & What questions are being asked depends on the discipline; a field of study can be defined by the set of questions it asks, the methods that are used to answer the questions, and other norms within the field \cite{hassan2022}. Clearly, the questions being asked, the methods being used, and the norms within a field evolve over time. & The question `What is Open Source Software?' has been addressed by several academic disciplines; SE research distinguishes itself from, say, Economics through the use of different theories and methods to answer the question. \\
    
    \addlinespace

    Paradigm & A paradigm is a way of looking at the world. Within scientific inquiry, it defines how researchers ask questions, and how those questions are answered. Paradigms imply a set of shared assumptions, and the questions that are worth asking.  & The Experimental Paradigm in Software Engineering  \cite{basili1993experimental}; the Evidence Based Software Engineering paradigm \cite{kitchenham2004evidence}. Software Engineering as a branch of engineering \cite{wasserman1996toward}; Software Engineering as a socio-technical field \cite{joblin2022successful}; Software architecture as a set of design decisions \cite{jansen2005software}\\

    \addlinespace

    Law &
    A law is a generalized statement that has universal applicability. Some laws are supported by both strong theory and empirical observations (e.g. Newton's universal law of gravity), others appear to be statements describing phenomena that appear to have universal application without a theoretical explanation.
    & Conway's Law, used to frame Herbsleb and Grinter's  study ``Conway's Law Revisited'' \cite{herbsleb1999splitting}; Lehman's Laws of Software Evolution; Hyrum's Law\footnote{https://www.hyrumslaw.com/}\\
    
    \addlinespace

    Framework &  
    A framework is an analytical tool that comprises the main concepts of interests, and potentially any relationships between them. Frameworks can serve many different purposes \cite{schwarz2007understanding}, and can be used to design a study, or be the outcome of a study.
    & Stol and Fitzgerald \cite{stol2014researching}'s research framework to study crowdsourcing, identifying six important themes within the domain of crowdsourcing when applied to SE.\\

\end{longtable}
\endgroup 

\begin{svgraybox}
\textbf{Exercises}\\
\begin{enumerate}
\item  Consider one of the following topics:\\[.1in]
Large Language Models, Secure Coding, Diversity and Inclusivity in SE\\

Identify a question to ask about this topic that is relevant to the software engineering community. 
Why is this question relevant to SE, and why is it important? What could be a question that is not relevant to SE?\\

\item Two popular paradigms within SE research are Evidence-Based Software Engineering \cite{kitchenham2004evidence} and Behavioral Software Engineering \cite{lenberg2015behavioral}. What are the assumptions underpinning each? What type of questions does each elicit? Suggest a question that is representative of each paradigm for the selected topic.\\

\item There are many `laws' within SE (Brooks' Law and Conway's Law as discussed above; Lehman's Laws of Evolution, etc.) \cite{endres2003handbook}. Identify a law within the SE domain, and how this could frame new research problems.\\

\item Given one of the topics listed above, perform a quick (ca. 1h), ad hoc literature review, and identify some themes that appear to be relevant to the picked topic. Create a table with these themes, and summarize for each theme why it is relevant to the topic at hand, and identify at least one question for each theme.

\end{enumerate}
\end{svgraybox}

\section{Theory Generators}
A second set of intermediate products of theorizing can help to \textit{generate} theory; these include myths, analogies, metaphors, and models \cite{hassan2022}. They help to generate theory because they can offer an initial foundation to capture some phenomenon of interest.  This section describes these (see Table~\ref{tab:generators} for a summary).

\subsection{Myth}\label{sec:myths}
A myth is an unquestioned belief that is not supported by evidence or fact \cite{trice1984studying}, or \textit{``A dramatic narrative of imagined events, usually used to explain origins or transformations of something''} \cite{hirschheim1991symbolism}.
Hassan et al. \cite{hassan2022} argued that myths are theory generators because \textit{``they can help uncover unquestioned assumptions within existing belief systems and theories.''} 
Myths help researchers to focus their research, to question whether the myth is true or not, and if not, pursuing an answer to the question what the truth then might be. 
Myths abound in SE research \cite{jorgensen2013myths}.
A classic example of this is the traditional waterfall approach to software development, described and critiqued by Royce \cite{royce1987managing}. In brief, the waterfall process describes a rational process, whereby first the requirements are gathered, followed by a system design, implementation, and then testing. While this approach appears to be completely rational in that \textit{it makes sense} to the outsider, anybody who ever developed any software knows this is not possible. Parnas and Clements \cite{parnas1986rational} discuss this mythical rational process, argue that it is an ``idealization,'' but then provide reasons as to why a description of such a process is useful. In effect, taking the myth of the rational development process, Parnas and Clements theorized why it is a myth, and continued to argue for faking it nevertheless.

Another example of this is a study by Rolland et al. \cite{rolland2023} (briefly mentioned in Sec.~\ref{sec:variance_process_theory}), who challenged three assumptions that were widely held in relation to the scaling of agile methods. In brief, while agile methods were originally meant for small projects performed by co-located teams, it soon became clear that organizations wanted to apply agile methods at scale. A considerably body of knowledge emerged over the past decade or so, reporting on such tactics like Scrum-of-Scrums to scale up agile methods. In their study, Rolland et al. identified three assumptions, or myths, that appeared quite dominant throughout the literature, for example: ``Agile and plan-driven methods are perceived to be mutually exclusive.'' This set the focus of the field study, through which they demonstrated that organizations need plan-driven practices to enable agility at scale. The analysis resulted in a series of propositions grounded in the empirical findings. Thus, the myth helped to \textit{generate} new theory (see Fig.~\ref{fig:processtheory}).

\subsection{Analogy}
An analogy can be used to explain or clarify something that is not (or less) familiar, by means of comparing to something that is known. In very simple terms, an analogy is saying X \textit{is like} Y, whereby X is the topic at hand, and Y is the analogy. 
While the two things differ in some ways, using an analogy means emphasizing the common aspects, finding similarities \cite{hassan2019process}. Analogies are useful to help explain a lesser known phenomenon or thing using terminology that is more familiar. 

As a simple example, consider the practice of pair programming. When we say that pair programming ``is like riding a tandem bicycle,'' we are using an analogy to convey that two people are working together to make progress on a task.\footnote{Lazy back-riders notwithstanding.} We're not suggesting that pair programming \textit{is} riding a tandem bike. Analogies are closely related to metaphors, discussed below; while all metaphors are analogies, not all analogies are also metaphors. Within pair programming, we rely on the metaphors `driver' and `navigator' to convey the meaning of the two roles engaged in pair programming.

An analogy is a very effective literary device to explain an otherwise difficult or unknown topic. 
To give another example: Component-Based Software Development (CBSD) was a popular approach in the nineties that attracted considerable attention from SE researchers. It can be explained by analogy as building a house from standardized, components that are pre-fabricated off-site \cite{kruchten2004putting}. The analogy helps in understanding that components are built `off-site' (i.e. by other teams or companies); that they have some standardized interface to connect them together, and that they exhibit certain pre-defined qualities. 
However, the analogy may also explain why CBSD as originally envisaged has not quite become common practice and faded as a topic of research:\footnote{This is arguably a pessimistic view, and depends on one's definition of `component.' In the nineties, standard component architectures such as Microsoft's COM (and later DCOM), and Java's JavaBeans appear not to have achieved the widespread adoption that had been expected. At the same time, if we treat open source libraries as components, software reuse has become an incredible success. Software ecosystems found in many popular languages (including NPM (Node.js, JavaScript), R, and Rust) offer fantastic environments to build new software products quickly from these components.} in building construction, components that don't quite fit well can be glued together with expanding foam, which is commonly used to seal gaps around cavities and is ideally suited to fill irregular shapes.
In software development, however, we don't have the equivalent of expanding foam: interfaces, data-formats and protocols must be strictly adhered to, otherwise the system as a whole doesn't work (cf. Garlan et al. \cite{garlan1995architectural}).
I would be remiss not to mention the Perl programming language, known as the ``Duct Tape of the Internet'' \cite{hall1999perl}, but this is more a metaphor than an analogy, because we don't \textit{compare} the Perl language to duct tape in a structured manner (see Table~\ref{tab:ducttape}). Metaphors are discussed next.

\begin{table}[!h]
    \centering
    \caption{Nobody makes a comparison like this to explain what the Perl language is, because duct tape is used a figure of speech. For the duct tape metaphor to work, it is important that people know what it is, else it is not a good metaphor.}
    \label{tab:ducttape}
    \begin{tabular}{p{2cm}p{2.8cm}p{6cm}}
    \toprule 
    Attribute & Duct Tape & Perl language\\
    \midrule 
     Stickiness    & Ties things together & Ties different programs together.\\
     \addlinespace
     Versatility & Easy to tear off a piece to use quickly; easy to use on any surface & Easy to fit in a Perl script into any operating system environment, including practically all Unix and Linux variants, Windows, and MacOS.\\
     \addlinespace
     Maintainability & May not last long & Perl scripts sometimes seen as unmaintainable due to the Perl syntax that doesn't suit everybody, and which some have described as ``line noise'' and Perl as a ``write-only language.''\\
    \bottomrule
    \end{tabular}
 
\end{table}

\subsection{Metaphor}

Metaphors are similar to analogies, but whereas an analogy is stating that X is \textit{like} Y, a metaphor is stating that X \textit{is a} Y.\footnote{This is a simplification, and does not necessarily imply that the decision whether something is a metaphor or analogy depends on the presence of the words ``is like.''} The difference is small and sometimes subtle, also because all metaphors are analogies, but not all analogies are metaphors. A few examples follow.

When we use the term `publication pipeline,' we are using a metaphor to convey the process of producing research papers. Implied here is that a number of papers are `in production' in various states of readiness. What we don't mean is that we have an actual pipeline in our labs into which we pour research data, and which produces papers as output.\footnote{Notwithstanding the recent developments of large language models.} When we say that we don't have any `bandwidth' to take on more research projects, we are using the term metaphorically to convey we are at capacity. 
To stick with the domain of publishing our research, we can treat our academic publications as a `conversation,' whereby each publication contributes to that conversation \cite{berente2024}. Following this metaphor, it would only make sense to contribute to a conversation if you first listen carefully to others---that is expected in academic papers through a related work section. Thus, this metaphor helps us to theorize about how academic publishing works, or should work. As discussed in Sec.~\ref{sec:challenges}, the SE literature does not much resemble a conversation.

Metaphors are extensively used in the SE field \cite{kendall1993metaphors}.  One that has received considerable attention over a time-span of more than two decades is the term `Technical Debt.' The metaphor was first used by Cunningham \cite{cunningham1992wycash} to describe the danger of shipping ``not quite right'' code, and the trade-off that is made in order to be able to ship a product fast:

\begin{quote}
``Shipping first time code is like going into debt. A little debt speeds development so long as it is paid back promptly with a rewrite. Objects make the cost of this transaction tolerable. The danger occurs when the debt is not repaid. Every minute spent on not-quite-right code counts as interest on that debt. Entire engineering organizations can be brought to a stand-still under the debt load of an unconsolidated implementation, object-oriented or otherwise.''
\end{quote}

Kruchten \cite{kruchten2012technical} explicitly recognized the role of this metaphor as a theory generator in their editorial \textit{``Technical Debt: From Metaphor to Theory and Practice.''} Lamenting that any issues in software development could be labeled some sort of `debt', Kruchten et al. argued that: 
\begin{quote}
    ``We need a better definition of what constitutes technical debt and some perspective or viewpoints that let us reason across a wide range of technical debt. In short, we need a theoretical foundation.''
\end{quote}

\subsection{Model}
Similar to the term `framework,' the term `model' also has different meanings in a SE research context. Whittle et al. \cite{whittle2013state} succinctly describe that 
\textit{``in software engineering, a model is an abstraction of a running system.''} This approach to software engineering is better known as Model Driven Engineering (MDE).\footnote{One particular subset of MDE is Model Driven Development, which seeks to generate code based on a model. In MDD, the model must be exactly right so that correct code can be generated---one could say that the model can be executed; generating and compiling the code are just intermediate steps. Because details now matter and must be included for the model to be executable, it is no longer an abstraction but just a different way for a programmer to express their intention (instead of code).  Therefore, the term `model' in MDD diverges from how the term is used in this chapter. In contrast, in research we create models to simplify the topic of interest so as to make sense of it, to reason about it, and to analyze it.}
Renowned theorist Robert Dubin equated the term `model' to mean `theory,' though acknowledged that not everybody agrees with this view \cite{dubin1978theory}. In this chapter, following Hassan et al. \cite{hassan2022}, I use the term `model' not to refer to a theory, but as an intermediate product of theorizing.

Consistent with Whittle et al.'s description, the term `model' usually implies that it represents an abstraction from something in the real world, and in the context of this chapter that is how I use the term model. Abstraction implies simplification and the leaving out of details. Curran \cite{curran787}, discussing statistical modeling, describes the importance of simplification:

\begin{quote}
''If I build a model 787 in my backyard, and then fly 382 people to Charles de Gaulle, it's not a model 787, it is a freakin' 787. It  is the simplification that is the point.''
\end{quote}

Statistical models are common in SE research, in particular multiple regression analyses. Less popular, though more powerful and flexible, are structural equation models (SEM) which are statistical models that can be easily visualized as box-arrow diagrams, which are a common way to represent variance theories (see Sec.~\ref{sec:variance_process_theory}). Statistical models have an associated model \textit{fit}; in multiple regression this is \textit{R}\textsuperscript{2}, indicating the variance explained by the fitted regression line, and in SEM there are a whole range of fit indicators \cite{stol2021gamification}. These measures of fit indicate how well the theorized model `fits' the data sample, and acknowledge that models are, in the end, simplifications that leave out details. In practice this means not only that variables are left out of a model (thus reducing the \textit{R}\textsuperscript{2}), but also the fact that a quantitative analysis cannot capture the idiosyncratic, and the specific details that could provide further insight.

A well-known statistical model in SE is Boehm's COCOMO cost estimation model \cite{boehm1984software}. 
But, not all models are statistical.  Redwine and Riddle \cite{redwine1985}'s technology maturity model (even though they didn't call it a `model') describes how technologies mature and get adopted over time, which can help to predict and set expectations.  
Models are used extensively in the literature on software processes \cite{rombach1995}, which describe processes, though models could describe other things as well. The waterfall approach described earlier (see Sec.~\ref{sec:myths}), which was documented and criticized by Royce \cite{royce1987managing}, is a model of the lifecycle of software development. 
The utility of such models like those described by Redwine and Riddle \cite{redwine1985} and the waterfall model can be assessed by looking at how closely they represent reality---how well these models `fit' with reality. As it turns out, the waterfall model is a rather poor fit with reality, in that most software development doesn't actually follow such a linear process.\footnote{Software development in safety-critical domains, such as the aerospace domain, automotive, and medical devices, is subject to regulatory guidelines or standards that prescribe the use of plan-driven methods.} 

Models provide the researcher to describe something in a systematic way so as to communicate the important characteristics, leaving out the idiosyncratic issues from the real world. As an abstract representation, researchers can reason about it, perform analyses, and even run computer simulations which themselves rely on simplifications \cite{stol2018abc}. Simulations are discussed in more detail in De Fran\c{c}a's chapter in this volume.

\begingroup
\footnotesize 
\begin{longtable}{p{1.9cm}p{4.9cm}p{4cm}}
\caption{Theory generators}\label{tab:generators}\\
    \toprule
    Product & Description & Example in SE\\
    \midrule
\endfirsthead
    \toprule 
    Product & Description & Example in SE\\
    \midrule
\endhead
   \bottomrule
\endfoot
    \bottomrule     
\endlastfoot
            
    Myth &     
    A myth is an unquestioned belief that is not supported by evidence or facts \cite{trice1984studying}. Challenging a myth can help to identify a new research direction.
    & The assumption that agile methods and plan-driven methods are mutually exclusive is a myth that Rolland et al. \cite{rolland2023} challenged so as to generate new theoretical insights. \\ 

    \addlinespace
    
    Analogy &     
    An analogy is a comparison of a phenomenon of interest to a phenomenon that is more familiar. The goal is to draw parallels so as to explain what something is, or how or why something works. 
    & A software system grows like a city; that is, a software system isn't a city, but it evolves  \textit{like} a city: it is in a constant state of evolution, changing some parts. \\

    \addlinespace

    Metaphor & 
    A metaphor is a figure of speech that equates two things, thereby assigning characteristics of the one thing to the other. All metaphors are analogies, but not all analogies are metaphors. 
    & Software Engineering; software development pipeline; software ecosystem; scaffolding code \cite{brooks1987essence}; Technical Debt \cite{kruchten2012technical}. \\

    \addlinespace
    
    Model & 
    
    A model is a simplified representation of a phenomenon that includes only those aspects that appear relevant and important, disregarding any details that may be idiosyncratic or incidental. A model helps to reason about the phenomenon of interest. The more detail a model has, the more realistic it is likely to be, at the cost of increased complexity.
    & Statistical models including structural equation models; software process models, including waterfall as a simplification of the software development life-cycle; computer simulations that are configured through a fixed number of parameters. \\

\end{longtable}
\endgroup

\begin{svgraybox}
\textbf{Exercises}\\
\begin{enumerate}
    
    \item Consider your own research topic. Attempt to find a paper that presents a ``model'' of in relation to the topic, and also a paper that presents a ``framework'' for the same topic. How are they used? How do they differ? \\
    \item Discuss the difference between an analogy and a metaphor. How do they differ? \\
    \item Identify a myth in the software engineering literature, and find one or two papers that discuss or investigate it. How could this myth be used as a theory generator?\\

    \item Identify a `law' in the software engineering literature, and find one or two papers that discuss or investigate it (cf. \cite{endres2003handbook,sachslaws}). Well-known examples are Brooks' Law, Conway's Law, Hyrum's Law. How does the law help to frame research?
\end{enumerate}
\end{svgraybox}

\section{Components of Theory}

The third set of theorizing products are actual components, or elements, of theory. These include the things we wish to reason about or measure (concepts, constructs), and the relationships between those things (statements, hypotheses). These components are commonly explained in a tutorial on `theory' \cite{stol2015theory}.
Because the terms concept and construct are so closely related, in defining one I make reference to the other. Table~\ref{tab:components} presents a summary.

\subsection{Concept}\label{sec:concepts}

Concepts are \textit{``ideas that receive names''} \cite{hassan2022,thomson1961}, and as such, they are the building blocks to develop theories. A research discipline deliberates the things that scholars believe are important, who then go on to name them. They have been described as the \textit{``product of the imagination that can be conveyed to others only by means of language''} \cite{jackendoff1989concept}.

The term `concept' is frequently conflated with the term `construct' (discussed next in Sec.~\ref{sec:constructs}); however, it can be useful to distinguish these as different things. Concepts represent sets of ideas, which are labeled to discuss them or theorize about them. Constructs, in contrast, are used to bring concepts to life, to operationalize them through some form of measurement. Hassan et al. \cite{hassan2022} summarize that \textit{``constructs are attempts to make concepts less `abstract' even though constructs are abstractions of concepts.''} 

Not all concepts are measurable, and therefore it is not always possible to define meaningful constructs for a concept. A concept can be any ``idea that receives a name,'' which also includes `software development process' or `software product,' and yet we would not seek to measure these concepts. Instead, we would measure their \textit{properties} \cite[p. 40]{dubin1978theory}; for example, rather than measuring the software product, we'd seek to measure its various qualities. In other words, concepts can be much more generic ideas.
Constructs are discussed in more detail below (Sec.~\ref{sec:constructs}); that section also describes one historical reason as to why the terms concept and construct may be conflated, or even considered equivalent by some. 

Of all intermediate theorizing products, concepts are perhaps the most straightforward: anything that has been labeled as something worthy of study. Some examples now follow.
At the time of writing, a current topic attracting attention within SE is `Developer Experience' (DevEx) \cite{fagerholm2012developer}.
The DevEx concept was first defined by Fagerholm and M\"{u}nch \cite{fagerholm2012developer} as: \textit{``how developers feel about, think about, and value their work.''}
However, the question then arises as to how to operationalize this definition, and this would require us to develop one, or perhaps more \textit{constructs}.

Another example of a concept that originated in the social sciences is the concept \textit{social capital}, which has also been used in SE research \cite{qiu2019going,stol2024jsis}.
Social capital has been defined as: \textit{``the goodwill that is engendered by the fabric of social relations and that can be mobilized to facilitate action''} \cite{adler2002social}. 
Nahapiet and Ghoshal \cite{nahapiet1998social} argued that social capital has three dimensions: the structural dimension, which includes network ties; the relational dimension, including trust and norms, and the cognitive dimension, which includes shared language. Each of these, social capital, network ties, trust, norms, shared language, are concepts, because they have a place in theoretical reasoning. We can define, teach, and reason about these concepts. To measure these concepts quantitatively, however, we must define constructs, discussed next.

\subsection{Construct}\label{sec:constructs}

Constructs are, in short, mechanisms to allow measurement (observation) of concepts.  Ironically, `construct' is the term frequently used to refer to variables that cannot be observed directly. Whereas concepts are the entities that we theorize about, the term construct becomes relevant when we attempt to measure those constructs. 
The terms construct and concept are frequently conflated; some scholars appear to reject any distinction \cite{rigdon2012rethinking}. Suddaby \cite{suddaby2010editor} acknowledges that the term `construct' comes with \textit{``connotations of hypothesis testing and operationalization,''} and that for researchers following a different research paradigm (see footnote \ref{foot:paradigm} in Sec.~\ref{sec:paradigm}), ``\textit{concept} might be a more acceptable value-neutral term.'' Whether one wishes to draw a distinction or not between `construct' and `concept,' it is important to achieve ``construct clarity'' (or conceptual clarity), i.e., clearly defined constructs (or concepts) \cite{suddaby2010editor}.

When we adopt a quantitative research strategy, it is useful to draw the distinction between ``construct'' and ``concept''; to clarify why, I make a brief, though technical, diversion.
I do this by drawing on Rigdon's explanation of what Bagozzi and Philips called the `Holistic Construal' \cite{bagozzi1982representing,rigdon2012rethinking}.

\begin{figure}[!h]
    \centering
    \includegraphics[width=\textwidth]{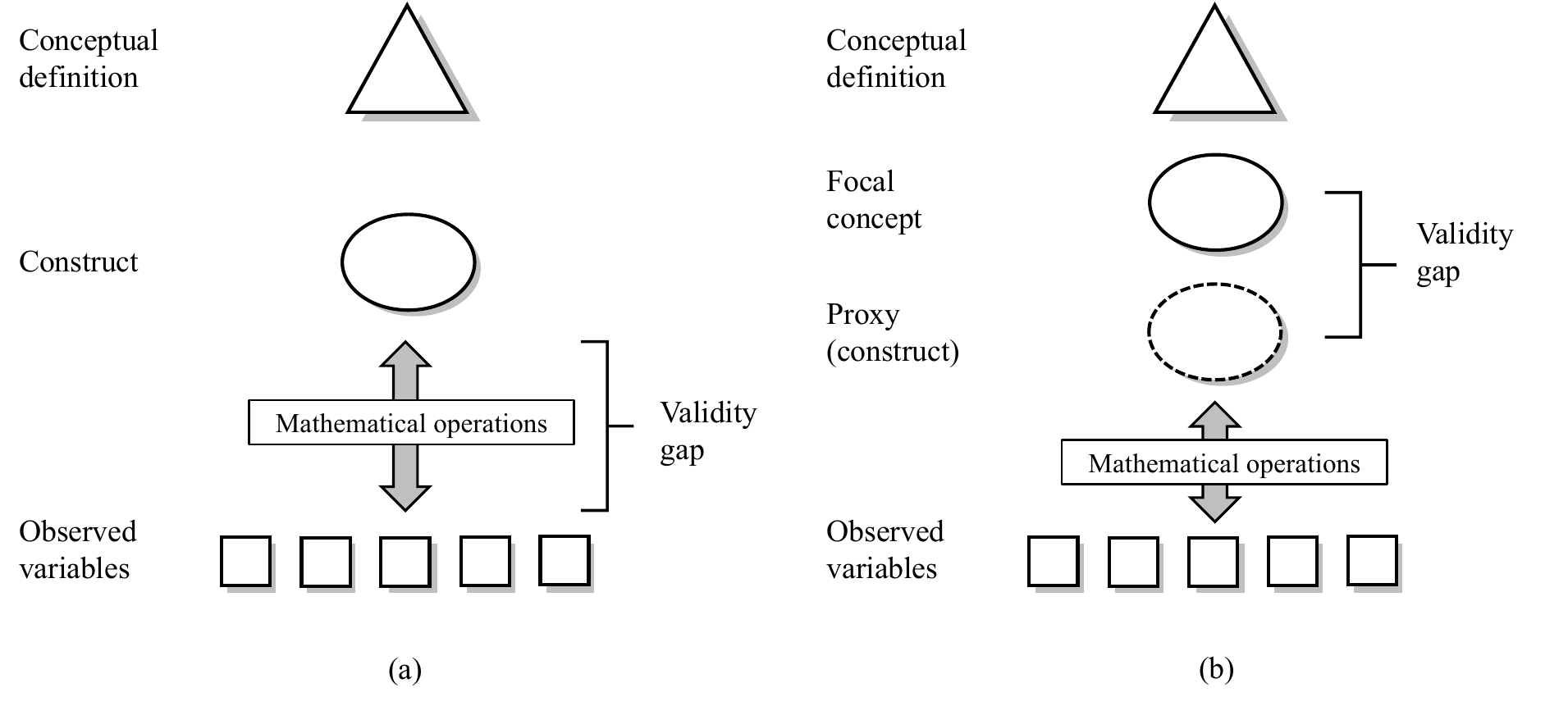}
    \caption{(a) Simplified representation of the Holistic Construal Framework \cite{bagozzi1982representing}, adapted from Rigdon \cite{rigdon2012rethinking}. (b) Concept proxy framework, adapted from Rigdon \cite{rigdon2012rethinking} that  explicitly acknowledges a distinction between concept and construct (proxy).}
    \label{fig:construct}
\end{figure}

Figure~\ref{fig:construct}(a) presents a simplified version of the Holistic Construal Framework (adapted from Rigdon \cite{rigdon2012rethinking}), which was first proposed in 1982, at a time when statistical techniques were not yet as well developed as today. The goal of the framework was to link together \textit{theory formulation} at the level of theoretical concepts, and \textit{theory testing} through constructs that represent those concepts. 

\begin{figure}[!h]
    \centering
    \includegraphics[width=4in]{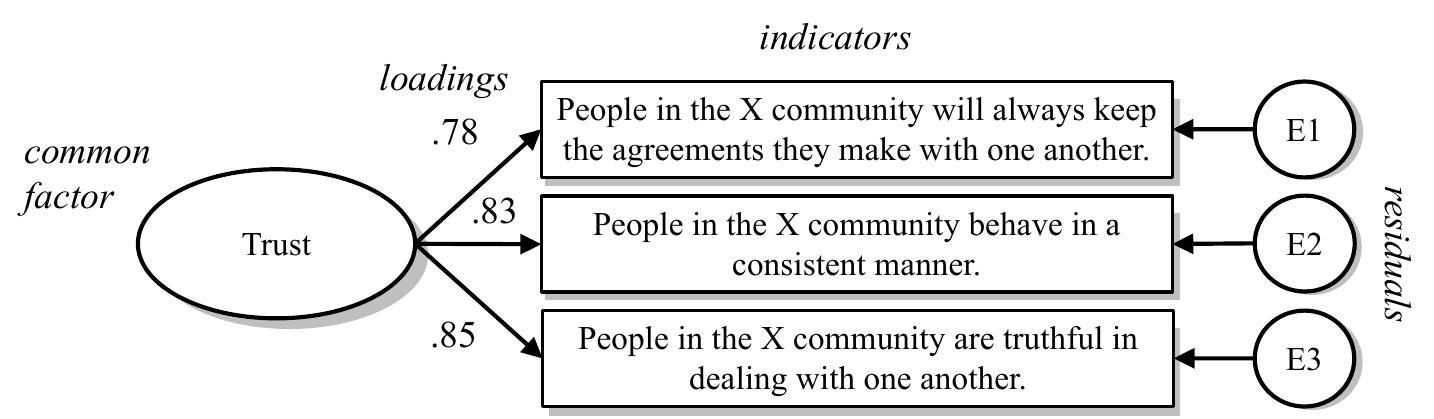}
    \caption{Trust as a reflectively measured common factor: instead of probing respondents on `trust,' which could involve misinterpretation and thus measurement error, we ask respondents to rate a set of more specific, unambiguous indicators, each of which has a certain loading on the common factor, indicating how much of its variance is explained by the common factor. Residuals explain the remaining variance in the indicators. X is a placeholder for the name of a community, not the social media platform formerly known as Twitter.}
    \label{fig:trust}    
\end{figure}

Research involves theoretical concepts; in behavioral software engineering, such concepts include job satisfaction, trust, personality traits, and a wide range of emotions. These concepts can be clearly defined to convey to readers what the researcher means, to achieve conceptual clarity \cite{suddaby2010editor}. To test a theory involving these concepts, one has to \textit{measure} these; to that end, a researcher would define \textit{constructs}, which are created through some mathematical operation from a set of observed (empirical) variables.  At the time the Holistic Construal Framework was first proposed, the dominant way to represent constructs was as reflectively measured \textit{common factors} (see also the chapter by Ralph et al., this volume).
Figure~\ref{fig:trust} shows an example for the construct `trust' (as used in Stol et al. \cite{stol2024jsis}) using standard notation in SEM: ovals represent latent variables, and empirical indicators are represented by rectangles. Reflective measurement refers to the idea that a theoretical, latent variable (e.g., trust) cannot be measured directly, but only \textit{inferred} through a set of indicators. Change in those indicators is \textit{caused} by change in the latent variable, i.e., change in the latent variable is  \textit{reflected} in the indicators. The relationship between a common factor and its indicators is expressed as a \textit{loading}, which indicates how much of the variance in the indicator is explained (`caused') by the common factor. Normally, not all variance in the indicators is explained by the common factor; the remainder is called the \textit{residual}, and this accounts for measurement error.

The apparent `natural fit' between theoretical concepts on the one hand, and their measurement through constructs represented by common factors contributed to a ``blurring'' of the terms concept and construct \cite{rigdon2012rethinking}. The lack of distinction has practical implications, because this implies that constructs \textit{are equal to} the concepts they represent. Under this assumption, `construct validity' is then the extent to which the construct is properly constructed from observed variables.

However, Rigdon \cite{rigdon2012rethinking} has argued that the implications of equating concept with construct are untenable. Doing so would equate a concept with a common factor,  the error terms on the indicators would perfectly capture measurement error, and that somehow, whatever conceptual definition the researcher had in mind was \textit{exactly} captured by the common factor (i.e., the construct).
Rigdon argued that the common factor is merely one approximation of whatever focal concept a researcher may have in mind, and that other representations are possible. He proposed the Concept Proxy Framework as a modified version of the Holistic Construal Framework (Figure~\ref{fig:construct}(b)), which explicitly acknowledges a distinction between the focal concept as ``idealized and out of reach,'' and a \textit{proxy} that is an ``empirical stand-in for the theoretical concept'' \cite{rigdon2012rethinking}. Like the construct in the Holistic Construal, the proxy can be created through a mathematical operation. From this perspective, the proxy may be modeled as a common factor or as something else, for example, a composite (a weighted combination of indicators) such as those in Partial Least Squares (PLS) SEM \cite{rigdon2012rethinking,russo2021csur}. The question of validity then shifts to the extent to which the proxy is  sufficiently similar to the theoretical concept that it purports to represent.

In sum, the distinction between the terms concept and construct is useful, because in quantitative studies we must operationalize the concepts we study; we do so by means of constructs. Whether the construct is statistically modeled as a common factor, or as a composite depends on many factors beyond the scope of this chapter, but whichever mathematical operation is selected, it is important to draw a distinction between the concept as something that is clearly defined and accept that it cannot be perfectly measured, and the construct that is the `empirical stand-in' to represent that concept.

\subsection{Statement}
Statements are the fundamental building blocks of theorizing \cite{hassan2022}.
In this context, the term statement does not refer to just any sequence of grammatically ordered words \cite{hassan2022}. Instead, it should be seen as an expression that makes sense within a given scientific discipline, i.e. it is relevant to what Hassan et al. \cite{hassan2022}, citing Foucault \cite{foucault1972}, refers to as scientific communication that makes up a \textit{discourse}. For example, the statement ``Continuous delivery improves software quality'' makes only sense within the context of SE research.\footnote{Note this is an example statement, not necessarily a universal fact.} Statements are claims that may or may not be presented with evidence. It is up to the researcher to (a) identify or develop the statement, and then (b) offer evidence in support of that statement.

While concepts (Sec.~\ref{sec:concepts}) and constructs (Sec.~\ref{sec:constructs}) are building blocks that represent phenomena and things of interest, statements (and also hypotheses, discussed below) can be made to discuss those concepts and constructs in relation to one another. 

A special type of statement is a proposition, which can be described as a ``truth statement,'' a ``prediction'' \cite{dubin1978theory}, or assertions about concepts \cite{hassan2022}. Hassan et al. \cite{hassan2022} distinguish different types of propositions, depending on the type of assertion.
An \textit{existential} proposition is a nonrelational proposition that simply declares the existence of a concept. This also includes typologies and taxonomies \cite{hassan2022,gregor2006,ralph2018toward}. 
A \textit{definitional} proposition goes beyond claiming the mere existence of a concept, and unpacks the concept, providing a definition and characterization.  A good example of this is work by De Boer et al. \cite{deboer2007,deboer2008}.
They observed that, while many researchers discussed architectural knowledge, it remained  ``a fuzzy concept.'' Thus, while there was general agreement on the \textit{existence} of the concept `architectural knowledge,' they attempt to unpack \textit{what architectural knowledge is} by drawing on different definitions and identifying its constituent elements.

Propositions should be made based on ``logical links'' rather than just any unjustified claim \cite{hassan2022,stol2015theory}. This is not to say that it's not possible to present a proposition that challenges the current state of knowledge (or assumptions held within a field); rather, such a challenge should come with a \textit{rationale}, a line of argument, a reasoning. Thus, the researcher should offer a reasonable derivation that leads up to the proposition. At the same time, there is little point in offering trivial propositions that have already been demonstrated by others, or in other fields. 
While any number of propositions can be proposed, Dubin suggests to aim for parsimony by focusing on \textit{strategic propositions} \cite[p. 168]{dubin1978theory}; that is, ignore the trivial propositions that do not add much insight, and focus on those propositions \textit{``where something notable is happening''} \cite[p. 169]{dubin1978theory}.

\subsection{Hypothesis}
Propositions (discussed above) are statements at the conceptual or theoretical level, but in order to test them, concepts must be operationalized using empirical indicators (and/or constructs, see Sec.~\ref{sec:constructs}) \cite{stol2015theory,hassan2022}. The result is a hypothesis. When operationalizing a proposition, it is possible that multiple hypotheses are necessary; for example, when hypothesizing about program size, one may need to use multiple metrics to operationalize this concept. These could include lines of code, number of statements, number of classes (when using object-oriented languages), or memory footprint.

Sometimes the term `hypothesis' is used in a more casual way, as if it were offered at a cocktail party. For example, closely related to Conway's Law (discussed earlier) is the ``Mirroring Hypothesis,'' which can be stated as: \textit{``Loosely-coupled organizations will tend to develop products with more modular architectures than tightly-coupled organizations''} \cite{maccormack2012exploring,colfer2016mirroring}. The form of this statement is more like a proposition than a hypothesis, because it only references  concepts such as ``loosely-coupled organizations'' and ``modular architectures.'' To actually test this hypothesis, we'd need empirical indicators (constructs); how would one measure the modularity of an architecture? For example, as part of their study, MacCormack et al. \cite{maccormack2012exploring} operationalize a dependency between modules as a function call; if a module A calls a function in module B, then A depends on B.

\begingroup
\footnotesize 
\begin{longtable}{p{1.9cm}p{4.9cm}p{4cm}}
\caption{Theory components}\label{tab:components}\\
    \toprule
    Product & Description & Example in SE\\
    \midrule
\endfirsthead
    \toprule 
    Product & Description & Example in SE\\
    \midrule
\endhead
   \bottomrule
\endfoot
    \bottomrule     
\endlastfoot
                
    Concept & 
    A name given to an idea or set of ideas so that it can be reasoned about. &  
    Developer Experience (DevEx) first proposed as a concept by Fagerholm and M\"{u}nch \cite{fagerholm2012developer}, and further studied by others \cite{noda2023devex}.\\

    \addlinespace
    
    Construct & 
    A proxy that is an empirical stand-in for a theoretical concept \cite{rigdon2012rethinking}; these can be modeled as a common factor (when using covariance-based SEM), a composite (when using PLS-SEM), or other forms such as sum scores, averages, etc. In SE, it is not uncommon to use singular observed variables as a proxy for a concept.
    &  Trust \cite{stol2021gamification}; social capital \cite{stol2024jsis}, and social capital \cite{qiu2019going} \\

    \addlinespace
   
    Statement & 
    A description relating to one or more concepts that is derived from logical reasoning, prior theoretical insights or literature, or empirical evidence. Includes propositions, typologies, and taxonomies. 
    & Statements include propositions (e.g. Crowston's propositions on work practices in open source software \cite{crowston2004effective}), and typologies and taxonomies \cite{ralph2018toward} (e.g. Lehman's SPE classification \cite{Lehman1980}.) \\

    \addlinespace
    
    Hypothesis & 
    A statement that has been operationalized so that it can be empirically tested.
    & Many papers in the top conference in SE feature hypotheses; specific examples can be found in \cite{stol2024jsis}, e.g.: The more a developer trusts other members in the inner source community, the higher that developer’s job satisfaction.  \\

\end{longtable}
\endgroup 

\begin{svgraybox}
\textbf{Exercises}\\
\begin{enumerate}
    
    \item Discuss the difference between a construct and a concept. How do they differ?\\
    \item Discuss the difference between a statement and a hypothesis. How do they differ?\\
    \item Consider a topic (phenomenon) of interest, for example, your current research topic. What are some of the key concepts that appear to be important that have received attention in prior work on this topic? Likewise, are there any defined constructs? \\
    \item Continue with the same topic; what are some key relationships that appear to be important? What are some well-known assertions, propositions, or hypotheses that appear in the relevant literature? 
    
\end{enumerate}
\end{svgraybox}

\section{Theorizing Products in Action: The Case of Software Architecture}\label{sec:map}

\begin{figure}
    \centering
    \includegraphics[width=\textwidth]{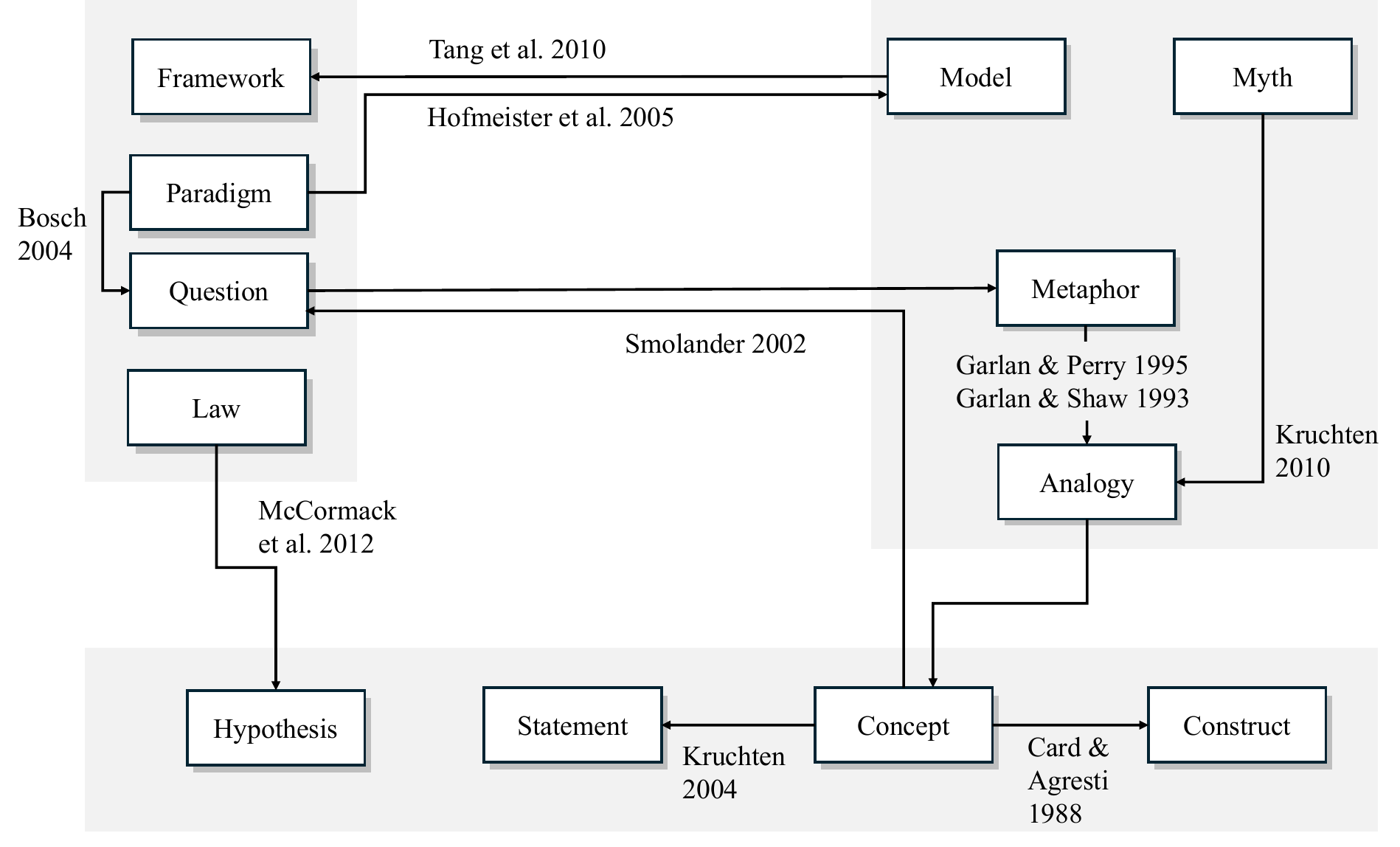}
    \caption{Example map of products of theorizing in the study of software architecture. This map is only one illustrative example of the author's (possibly naive) interpretation of how studies relate to one another. The author acknowledges this map is incomplete. Further, in creating a map, any theorizing product can be linked to any other. Creating a more complete map can be a useful yet time-consuming way to learn about a research area}
    \label{fig:map}
\end{figure}

The theorizing products presented above are drawn from Information Systems scholars \cite{hassan2022}. One could question, or even object, that such a theoretical focus is not applicable to a practical field like SE, a field that some perceive as an engineering field. To highlight the value of these theorizing products, I now demonstrate how theorizing takes place within the software architecture field, an area of research that is firmly rooted within the SE domain. I adopt Hassan et al.'s approach \cite{hassan2022} to illustrate how studies move the field forward by leveraging one theorizing product, to move to another theorizing product (Fig.~\ref{fig:map}). The presentation below should not be construed as a complete and necessarily accurate presentation of how theorizing \textit{actually} took place. Further, while the various studies all relate to the software architecture field, the studies used to illustrate the method link to different topics, including software architecture as a concept and architectural knowledge. 
Reflecting on how an area of research evolved using the 12 theorizing products as a lens can help to increase awareness of theorizing so that future researchers can also learn how to reflect on a field.

\subsection{From Law to Hypothesis}
Among the early literature concerned with the structure and decomposition of software, years before the term software architecture was coined, is an essay by Conway who suggested that the structure of a system will follow the communication structures that exist within the organization that produces that system \cite{conway1968committees}. This has been labeled Conway's Law in the context of software systems, but similar observations were made by others, and this has become more generally known as the ``mirroring hypothesis.'' MacCormack et al. phrased this as \cite{maccormack2012exploring}:
\textit{``Loosely-coupled organizations will tend to develop products with more modular architectures than tightly-coupled organizations.''}

\subsection{From Metaphor to Analogy to Concept}
The software architecture sub-field has deep roots dating back to the very early days of software engineering, but started gaining traction in earnest in the mid to late 1980s. Garlan and Perry \cite{garlan1995introduction} noted that one of the trends that prompted the attention for software architecture:

\begin{quote}
For example, the box and line diagrams and explanatory prose that typically accompany a high-level system description often refer to such organizations as a ``pipeline,'' a ``blackboard-oriented design,'' or a ``client-server system.''

\end{quote}
Pipeline, blackboard, and client-server are all metaphors that convey a particular architectural style, the overall structure of a system. There was an increasing recognition that there is such a thing as software design at the system level, rather than at the detailed level. Up to that point, detailed design would have been captured using dataflow diagrams \cite{yourdon1989}, for example. 
One of the fundamental papers introduced software architecture as follows \cite{garlan1993introduction}: 

\begin{quote}
    As the size and complexity of software systems increases, the design problem goes beyond the algorithms and data structures of the computation: designing and specifying the overall system structure emerges as a new kind of problem. This is the software architecture level of design.''
\end{quote}

Thus, an analogy was drawn to the high-level structure of a system, as suggested by the metaphors representing the architectural styles, similar to the idea that one can observe different architectural styles in buildings throughout the ages, and across the world. The term `architecture' is a metaphor\footnote{`Architecture' is the metaphor, not the term `software architecture,' because prior to the coining of the term, it didn't exist. The same applies to `technical debt': debt is the metaphor.} for high-level design of a software system, and  quickly found its way through into software companies, for example, in the naming of the role software architect.\footnote{The  analogy between building systems and construction has not gone without critique \cite{baragry2001we}. I agree that the metaphor is not perfect: software architects would normally have a strong development background, whereas builders don't usually become architects. Another point where the analogy with construction breaks down is the fact that architects in construction do, in fact, focus on details, aesthetics, and elegance. For example, architects would design `hidden' gutters and downpipes as these essential elements would take away from the aesthetics of a building when visible.}  By now, the term `architecture' (in the context of software systems) has become so common, it is a `dead' metaphor \cite{giles2001missing}.

\subsection{Deriving Constructs from Concept}
Card and Agresti \cite{card1988measuring} suggest that \textit{``architectural design complexity derives from two sources: structural (or intermodule) complexity and local (or intramodule) complexity.''} In effect, Card and Agresti derived two constructs from the concept of architectural design complexity, namely the structure of the system as a whole, and the complexity of individual components; each of these constructs is measured using specific metrics. 
While the definition of constructs is not a theory, defining constructs is essential to make any `theory' (in the broadest sense of the word) of software architecture testable. 

\subsection{Exploring Concept by Metaphor}
While software architecture had emerged as a concept, mostly as a level of design that went beyond algorithms and data structures, the question `What is Software Architecture' did not have a clear answer. Whereas there have been numerous working definitions, Smolander \cite{smolander2002four} observed in 2002 that the meaning of the concept remained ambiguous and that academic and industry perspectives diverged \cite{bosch2000design}. Around that time, the Software Engineering Institute (SEI) maintained a list of close to a hundred different definitions of software architecture. Smolander investigated how software companies perceived software architecture, and identified four metaphors: architecture as a blueprint, architecture as literature, architecture as language, and architecture as decision, representing the different roles that software architecture plays in practice. The last definition, architecture as decision, forebode a paradigm shift, discussed next. 

\subsection{Shift in Paradigm Leads to Shift in Questions}
The software architecture as a topic of interest has emerged and evolved rapidly over the past few decades. Initially the focus was on technical artifacts, considering architecture as a collection of components and connectors that exhibited a certain structure and behavior (e.g., performance, reliability, and other quality attributes) \cite{capilla201610}. Researchers studied how to capture this using different views, using architecture design methods and architecture description languages. Bosch \cite{bosch2004software}, among others, pointed out that this view on software architecture lacks attention for capturing design decisions, and consequently the knowledge about `what and how' is not captured in the software architecture product. This focus on design decisions heralded a paradigm shift; rather than focusing on software architecture as the end-product, the focus was turned to the architectural design process, including capturing design decisions and managing architectural knowledge \cite{capilla201610}. 

As a result, researchers started asking different types of questions. Instead of asking how we can capture different architectural styles or patterns in UML, for example, researchers now were focusing on defining architectural knowledge \cite{deboer2007}, and how it could be managed.

This second paradigm of software architecture is not a theory, but helped to frame the research done in this field, the questions that researchers ask, and ultimately the theorizing products that come out of that inquiry.

\subsection{From Concept to Statement}
The `design decision' concept gained a lot of traction within the SA domain as it opened up a new view on what should be captured. Numerous templates have been proposed to capture various attributes of design decisions. Several researchers then started to study the concept of design decisions in more detail by not only proposing templates that capture the various attributes of a design decision, but also taxonomies and ontologies. For example, Kruchten \cite{kruchten2004ontology} presented an ontology of design decisions, that defines different types of design decisions, and different types of relationships between those decisions. In so doing, Kruchten theorized convincingly, using examples from an air traffic control system, that there are `existence' decisions (a system will have a certain element or artifact), `property' decisions (a system will have a certain enduring trait or quality), and `executive' decisions (decisions about the process, technology, or tools to be used). While the ontology is not a theory, it did help advance the SA field.\footnote{A crude measure of impact is that Kruchten's paper was cited 420 times at the time of writing.}

\subsection{From Model to Framework}
A variety of different architecture design methods emerged over time, mostly independently as they were created by different organizations in practice. Observing that these methods vary in terminology and focus, Hofmeister et al. \cite{hofmeister2005generalizing} distilled a general model of architecture design. This model very much aligned with the first paradigm of software architecture research, focusing on the software architecture as an end-product. Tang et al. \cite{tang2010comparative} extended this model to capture the full life-cycle of architecture, which includes architecture implementation and maintenance, in recognition of the fact that the architectural knowledge created in the early phases is used later, including when a system has been deployed and must be maintained. 

Whereas the extended model captures the full life-cycle of architecture design during which architectural knowledge (AK) can be created and used, Tang et al. \cite{tang2010comparative} then turned their focus on the activities that produce and consume AK. In so doing, they identify 12 different types of activities through which producers and consumers generate and use AK.

\subsection{Debunking Myth by Analogy}
In the earlier days during the rise of agile methods in industry, there was an apparent conflict between elements of the Agile Manifesto and software architecture as a `Big Up-Front Design' \cite{kruchten2010}. This myth is rooted in a false dichotomy between agile methods and architecture \cite{abrahamsson2010agility}, which could be the result from a superficial reading of the Agile Manifesto. Kruchten \cite{kruchten2010} approached to resolve this mythical conflict by drawing an analogy the ethno-sociological concept of culture and intercultural conflicts, emphasizing that the conflict can be reconciled by improving mutual understanding and foregoing \textit{``a shallow caricatural view of the `other culture.'\,''}

\section{Teaching Theorizing: A Practical Guide}
This chapter presents a wide range of theoretical concepts (pun intended), and in so doing,  scratches only the surface of each of the 12 intermediate theorizing products. Table~\ref{tab:guide} offers a practical guide to teaching this material; it should be noted that a 12-week course could focus on any of the 12 topics, or just a subset of these, informed by the lecturer's preferences. It is assumed there are 2 contact hours per week, for a total of 24 hours for the whole semester. This might seem overindulgent given there are so many other topics to teach, but this is, in the end, complicated material that requires considerable engagement lest the effort become meaningless. When faced with less time, rather than attempting to include too much material, I would suggest to focus on a select few themes, and discuss these in-depth.

Whatever the focus and scope is of one or a series of lectures or tutorials, it is important (a) to read one or two papers that set the stage (Table~\ref{tab:guide} offers useful starting points, but these recommended readings can easily be swapped for others); (b) to apply the material from the paper in one way or another. An effective question is: ``Given the reading assignment, what elements could be applied to \textit{my} situation or research interest?'' or ``what would this look like when applied to \textit{my} research topic?'' I have attempted to do this for the Software Architecture field in Section~\ref{sec:map}.\footnote{The astute reader will notice that the map has considerable room for improvement.}

All 12 topics are ``rabbit holes''\footnote{Question to the reader: a metaphor or analogy?}, i.e., it is easy to get lost. This is both the magic, but also the curse\footnote{Metaphor or analogy?} of doing meta-science; it is magical because there is so much more to learn, but it is a curse because there doesn't appear to be a final answer. Welcome to theorizing! 

\begingroup
\footnotesize
\begin{longtable}{p{1cm}p{1.5cm}p{8.4cm}}
\caption{Sample program for a 12-week course to teach theorizing in SE. Listed activities are examples only to get started, but lecturers are recommended to add or replace contents as they see fit.}\label{tab:guide}\\
    \toprule
        Week & Topic & Readings, Questions, Activities \\
    \midrule
\endfirsthead
    \toprule 
        Week & Topic & Readings, Questions, Activities \\
    \midrule
\endhead
   \bottomrule
\endfoot
    \bottomrule     
\endlastfoot
    Week 1 & Introduction; Metaphor \& Analogy & Read Weick on theorizing \cite{weick1995theory}. Find a few papers in your field of interest that use metaphors or analogies. Which analogies are also metaphors, and why? Which analogies are not, and why? \\
    \addlinespace
    Week 2 & Questions & Read Sandberg and Alvesson \cite{sandberg2011ways}. Identify a topic of interest, and identify two scientific disciplines (or sub-discplines) that study it, for example: SE, IS, CSCW/CHI, management, economics, psychology, etc. Attempt to identify questions asked in each discipline on this topic. \\
    \addlinespace
    Week 3 & Concepts & Identify a concept that has attracted interest in the past several years; examples include productivity, developer experience. What definitions are provided, if any?  \\
    \addlinespace
    Week 4 & Constructs & Read Suddaby \cite{suddaby2010editor}, Graziotin et al. \cite{graziotin2021psychometrics}, Sj\o{}berg and Bergersen \cite{sjoberg2022construct}, Ralph and Tempero  \cite{ralph2018construct}. Advanced reading is Rigdon's work on measurement, e.g. \cite{rigdon2012rethinking}. Discuss the differences between concepts and constructs; discuss construct validity; consider some concept within SE, and find an associated construct to measure it; is the construct a good representative of the concept?\\
    \addlinespace
    Week 5 & Frameworks & Read Leshem and Trafford \cite{leshem2007overlooking} and Schwarz et al. \cite{schwarz2007understanding}. Given a substantive research topic or question, attempt to create a (initial) framework to help answer this question.  \\
    \addlinespace
    Week 6 & Laws & Read Conway's short article \cite{conway1968committees}, and discuss the strength of evidence that Conway proposed; is the term `law' appropriate? Did later studies provide support? Identify some other `law' in SE: attempt to find the origin of the law, and papers that have cited the law to frame a study.  \\
    \addlinespace
    Week 7 & Myths & Identify a myth in SE, for example, see Bossavit \cite{bossavit2015leprechauns}. Attempt to identify the origin, and a paper that refutes the myth.  \\
    \addlinespace
    Week 8 & Models & Consider the model of architectural knowledge by Hofmeister et al. \cite{hofmeister2005generalizing} and the extension by Tang et al. \cite{tang2010comparative}; attempt to identify a model within a different field. Discuss the difference between a framework and a model. \\
    \addlinespace
    Week 9 & Paradigms & Read Basili \cite{basili1977software} and Kitchenham et al. \cite{kitchenham2004evidence}, and compare the two viewpoints; how do Basili and Kitchenham et al. converge or diverge? Consider a research topic/field, and attempt to identify trends over the past 20-30 years, e.g. sampling papers every 4-5 years; what appear to be assumptions within the field? Are there any `shifts' in focus when comparing earlier versus recent research? Is there a shift in methods used? Attempt to compare early vs. current papers using a table. Can you speak of a paradigm shift?  \\
    \addlinespace
    Week 10 & Statements \& Hypotheses & Read Crowston et al. \cite{crowston2004effective}. How did they develop the propositions from prior literature? Are these propositions `strategic' and `interesting'? If so, why; if not, why not? Read Ralph for more guidance on taxonomies \cite{ralph2018toward}.  \\
    \addlinespace
    Week 11-12 & Theorizing Map; wrap-up & Build a theorizing map (see Sec.~\ref{sec:map}, Figure~\ref{fig:map}) on a topic of choice, ideally linked to the same concepts. This is time-consuming and challenging, but is an excellent activity to explore historical developments within a field of interest.\\  
    
\end{longtable}
\endgroup 

\subsection*{Acknowledgments}
This work is supported with funding by Science Foundation grant 13/RC/2094-P2 to Lero, the SFI Research Centre for Software.

\bibliographystyle{spmpsci}

\end{document}